\documentclass[aps,twocolumn,superscriptaddress,showpacs,amsmath]{revtex4}

\usepackage{epsf,latexsym}
\usepackage{amssymb,amsmath}
\usepackage{times}

\newcommand{\bra}[1]{\langle #1 |}
\newcommand{\ket}[1]{| #1 \rangle}

\newcommand{\qed}{$\hfill \Box$}

\newcommand{\ignore}[1]{}

\newcommand{\be}{\begin{equation}}
\newcommand{\ee}{\end{equation}}

\def\CC{{\rm\kern.24em \vrule width.04em height1.46ex depth-.07ex
    \kern-.30em C}}
\def\P{{\rm I\kern-.25em P}}

\def\RR{{\rm
         \vrule width.04em height1.58ex depth-.0ex
         \kern-.04em R}}

\def\bbbone{{\mathchoice {\rm 1\mskip-4mu l} {\rm 1\mskip-4mu l}
{\rm 1\mskip-4.5mu l} {\rm 1\mskip-5mu l}}}

\def\bbbc{{\mathchoice {\setbox0=\hbox{$\displaystyle\rm C$}\hbox{\hbox
to0pt{\kern0.4\wd0\vrule height0.9\ht0\hss}\box0}}
{\setbox0=\hbox{$\textstyle\rm C$}\hbox{\hbox
to0pt{\kern0.4\wd0\vrule height0.9\ht0\hss}\box0}}
{\setbox0=\hbox{$\scriptstyle\rm C$}\hbox{\hbox
to0pt{\kern0.4\wd0\vrule height0.9\ht0\hss}\box0}}
{\setbox0=\hbox{$\scriptscriptstyle\rm C$}\hbox{\hbox
to0pt{\kern0.4\wd0\vrule height0.9\ht0\hss}\box0}}}}

\def\bbbz{{\mathchoice {\hbox{$\sf\textstyle Z\kern-0.4em Z$}}
{\hbox{$\sf\textstyle Z\kern-0.4em Z$}}
{\hbox{$\sf\scriptstyle Z\kern-0.3em Z$}}
{\hbox{$\sf\scriptscriptstyle Z\kern-0.2em Z$}}}}

\newcommand{\putfig}[2]{$$\leavevmode\hbox{\epsfxsize=#2 cm
   \epsffile{#1.eps}}$$}

\begin{document}

\title{Bipartite entanglement and entropic boundary law in lattice spin systems}

\author{Alioscia Hamma}
\affiliation{Institute for Scientific Interchange (ISI), Villa Gualino, Viale Settimio Severo 65, I-10133 Torino, Italy}
\affiliation{Dipartimento di Scienze Fisiche, Universit\`a Federico II, Via Cintia ed.~G, 80126 Napoli, Italy}

\author{Radu Ionicioiu}
\affiliation{Institute for Scientific Interchange (ISI), Villa Gualino, Viale Settimio Severo 65, I-10133 Torino, Italy}

\author{Paolo Zanardi}
\affiliation{Institute for Scientific Interchange (ISI), Villa Gualino, Viale Settimio Severo 65, I-10133 Torino, Italy}

\begin{abstract}
We investigate bipartite entanglement in spin-1/2 systems on a generic lattice. For states that are an equal superposition of elements of a group $G$ of spin flips acting on the fully polarized state $\ket{0}^{\otimes n}$, we find that the von Neumann entropy depends only on the boundary between the two subsystems $A$ and $B$. These states are stabilized by the group $G$. A physical realization of such states is given by the ground state manifold of the Kitaev's model on a Riemann surface of genus $\mathfrak{g}$. For a square lattice, we find that the entropy of entanglement is bounded from above and below by functions linear in the perimeter of the subsystem $A$ and is equal to the perimeter (up to an additive constant) when $A$ is convex. The entropy of entanglement is shown to be related to the topological order of this model. Finally, we find that some of the ground states are absolutely entangled, i.e., no partition has zero entanglement. We also provide several examples for the square lattice.
\end{abstract}


\maketitle

\section{Introduction}

Entanglement emerged recently as a quintessential concept in several fields of physics. In quantum information theory (QIT) entanglement is a {\em sine qua non} resource for various quantum processing and quantum communications protocols, like teleportation, dense coding, cryptography, and is crucial for the exponential speed-up of several quantum algorithms \cite{QC}. The same concept is also essential for our understanding of several solid-state systems. Examples are the entangled ground state for two highly non-classical systems: superconductivity (the BCS state \cite{zanardi,botero}) and fractional quantum Hall effect (the Laughlin state \cite{fqhe}). Another manifestation of the ubiquity of entanglement is found in the study of quantum phase transitions, where it is believed to be responsible for the appearance of long-range correlations \cite{phasetrans}.

Spin systems are also a distinguished playground for the study of bipartite entanglement and its scaling with the subsystem size. It has been shown that systems in which entanglement scales less than logarithmic can be efficiently simulated on a classical computer \cite{vidal}. Hence the amount of entanglement present in the system and its scaling is crucial for efficient quantum algorithms, i.e., problems that are classically intractable. In the case of a critical spin chain in  $XY$ and Heisenberg models, the entanglement between a spin block of size $L$ and the rest of the chain scales like $S\sim \log_2 L$ and thus this system can be simulated classically \cite{latorre}. Several groups have analyzed recently the entanglement properties of various spin systems \cite{cirac}, including 1-dimensional lattice models of the $XY$ \cite{latorre,korepin}, Heisenberg \cite{latorre} and Anderson model \cite{li}.

In this article we study bipartite entanglement in general spin systems extending our previous results found for the ground state of the Kitaev's model \cite{hiz}. The relevance of this model stems from the fact that it was the first example of the new subject of topological quantum computation \cite{kitaev,top} and because it features {\em topological order} \cite{wen}. This is a type of quantum order which describes states of matter that are not associated to symmetries, like the fractional quantum Hall liquids. We show that for a large class of states (e.g., states stabilized by  groups of spin flips, which include the ground state of the Kitaev model) the von Neumann entropy depends only on the degrees of freedom belonging to the {\em boundary} of the two subsystems. Hence, our result echoes the holographic principle \cite{holographic}: the geometric entropy of a region $A$ depends only on the degrees of freedom of the {\em boundary} of $A$, and not of the {\em bulk}.

The structure of the paper is the following. In Section II we expose the general formalism for bipartite entanglement in spin systems and we apply it to states that are stabilized by groups of spin flips. For spins on a lattice we provide a geometrical interpretation of these results. In Section III we exemplify this general framework for the Kitaev's model \cite{kitaev} and we apply it to calculate the ground state entanglement. We calculate analytically the entropy of entanglement for several partitions of the lattice, like spin chains and spin ladders in Section IV. We then conclude in Section V.

\section{Entanglement in a spin system}

\subsection{The reduced density matrix}\label{formalism}

In this section we find a general expression for the reduced density matrix of an arbitrary spin system and we give a necessary and sufficient condition for its diagonality. These results specialize in a very interesting way to those states, like the Kitaev's ground states $\ket{\xi_{ij}}$ (see Section \ref{kitmod}) that can be written as an equal superposition of all the elements in a group acting on the reference state $\ket{0}$. To begin with we do not assume any particular geometry or dimensionality of the spin system.

Given a system of $n$ spins-1/2, its Hilbert space has the tensor product structure $\mathcal H= \mathcal{H}_1^{\otimes n}$, where $\mathcal{H}_1= \mbox{span} \{ \ket{0}, \ket{1}\}$ is the Hilbert space of a single spin. In the usual computational basis we define a reference basis vector
\be
\ket{0} \equiv \ket{0}_1\otimes ...\otimes\ket{0}_n
\label{zero}
\ee
i.e., all spins up. Let $N= N_1^{\otimes n}$ be the Abelian group of all spin flips, where $N_1= \{ \bbbone, \sigma^x \}$ acts on a single spin. Obviously $\dim {\mathcal H}= |N|= 2^n$ and any vector of the computational basis can be written as $\ket{i}= g\ket{0}$, for some $g \in N$ and $i=0 \ldots 2^n-1$ (as a binary expansion). Moreover, all elements satisfy $g^2= \bbbone$, $g\in N$. A generic state in the Hilbert space can be written as the superposition of all the possible spin flips on this system, namely
\begin{equation}
\ket{\psi}= \sum_{g\in N}a(g)g \ket{0}
\end{equation}
with $a(g)\in \CC$, $\sum_g |a(g)|^2= 1$.

Consider now the states $\ket{\psi}$ that can be written as a superposition of elements obtained only from a subgroup of spin flips $G\subseteq N$ acting on $\ket{0}$. Then the corresponding density matrix is
\begin{eqnarray}
\nonumber
\rho &=& \ket{\psi}\bra{\psi}= \sum_{g,g'\in G\subseteq N} a(g)\overline{a}(g')g\ket{0}\bra{0}g' \\
&=&  \sum_{g,\tilde{g}\in G\subseteq N}a(g)\overline{a}(g\tilde{g})g\ket{0}\bra{0}g\tilde{g}
\end{eqnarray}
using the substitution $g'= g\tilde{g}$ in the last equation.

We now compute the reduced density matrix of an arbitrary subsystem $A$ of spins, hence tracing out over all the spins in the complement subsystem $B$. Any element $g\in G$ has a tensor product structure $g=g_A\otimes g_B$, with $g_{A,B}\in N$ acting only on $A$ and respectively $B$ subsystems. It is important to note that in general $g_{A,B}\not\in G$. Writing $\ket{0}\equiv \ket{0_A}\ket{0_B}$, we obtain
\begin{eqnarray}
\nonumber
\rho_A&=& \sum_{g,\tilde{g}\in G\subseteq N}a(g)\overline{a}(g\tilde{g})g_A\ket{0_A}\bra{0_A}g_A\tilde{g}_A\times\\
&\times& \bra{0_B}g_B\tilde{g}_Bg_B\ket{0_B}
\end{eqnarray}
where $\tilde{g}= \tilde{g}_A\otimes \tilde{g}_B$.

We introduce now two subgroups of $G$ acting trivially on the subsystems $A$ and respectively $B$:
\begin{eqnarray}
G_A &\equiv& \{g\in G\ |\ \  g= g_A\otimes \bbbone_B\} \\
G_B &\equiv& \{g\in G\ |\ \  g= \bbbone_A\otimes g_B\}
\end{eqnarray}
We denote their order by $d_{A,B}\equiv |G_{A,B}|$. With these notations, the only non-zero elements in $\bra{0_B}g_B\tilde{g}_Bg_B\ket{0_B}$ satisfy $\tilde{g}_B=\bbbone_B$ (since $g_B^2=\bbbone_B$) and this implies that $\tilde{g}\in G_A$. We finally obtain
\begin{equation}
\label{rhoa}
\rho_A= \sum_{g\in G,\tilde{g}\in G_A} a(g)\overline{a}(g\tilde{g}) g_A \ket{0_A}\bra{0_A} g_A\tilde{g}_A
\end{equation}

In general $\rho_A$ will contain off-diagonal terms. The following lemma gives the necessary and sufficient conditions to have a diagonal $\rho_A$ (in the computational basis).

{\em Lemma 1.} The following statements are equivalent:\\
(a) $\rho_A$ is diagonal; \\
(b) no $g\in G$, $g\ne\bbbone$ acts trivially on $B$, i.e., $G_A=\{\bbbone\}$; \\
(c) no element $g$ in $G$ can be decomposed as the product $g=g_1\cdot g_2$ with both $g_1, g_2$ nontrivial and in $G_A,G_B$ respectively.\\

{\em Proof.} $(a)\Leftrightarrow(b)$: from equation (\ref{rhoa}) $\rho_A$ is diagonal iff $\tilde{g}_A=\bbbone_A$, hence $G_A= \{\bbbone\}$ and $d_A=1$.

$(b)\Rightarrow(c)$: Since $G_A$ contains only the identity, there is no $g_1=g_A\otimes\bbbone_B$ different from the identity and thus there is no $g=g_1\cdot g_2$ with nontrivial $g_{1,2}\in G_{A,B}$, which proves the sufficient condition. We prove the necessary condition $(b)\Leftarrow(c)$ ex absurdo. If there were a nontrivial $g\in G$ such that $g=g_A\otimes\bbbone_B$, then we can write $g=g_1\cdot g_2$ with $g_1=g\in G_A$ and $g_2=\bbbone \in G_B$, contradicting the hypothesis. \qed

\subsection{Entropy of entanglement for a stabilized space}

We are interested to quantify the entanglement present in our spin system. Although there is no known entanglement measure for a general multi-qubit system, we can study bipartite entanglement of a system described by a pure density matrix $\rho_{AB}$. In this case the von Neumann entropy $S$ is the unique measure of bipartite entanglement:
\begin{equation}
S\equiv -\mbox{Tr} (\rho_A \log_2 \rho_A)
\label{entropy}
\end{equation}
where $\rho_A= \mbox{Tr}_B (\rho_{AB})$ is the reduced density matrix of the sub-system $A$. The von Neumann entropy of a density matrix $\rho$ is bounded by $0\le S \le \log_2 d$, where $d$ is the dimension of the Hilbert space of $\rho$. The bound is saturated iff $\rho=\bbbone/d$, i.e., the system is in the totally mixed state. For a bipartite system $(A,B)$ of $n$ spins we can readily obtain a simple bound for the entropy (using the symmetry $S= -\mbox{Tr}_B(\rho_A\log_2 \rho_A)= -\mbox{Tr}_A(\rho_B\log_2 \rho_B)$):
\be
0\le S\le \min(n_A, n- n_A)
\ee
where $n_A, n-n_A$ are the number of spins in the $A$ and $B$ partition, respectively.

We now apply the formalism of Section \ref{formalism} to states of a {\em stabilized space}. Let $\{ U_s \}$ be a set of mutually commuting operators, called {\em stabilizer operators}. A state $\ket{\psi}\in {\mathcal H}$ is {\em stabilized} if is invariant under the action of the stabilizer operators: $U_s \ket{\psi}= \ket{\psi}$, $\forall s$. Let $G$ be the group generated by the stabilizer operators $U_s$. The space $\mathcal L$ stabilized by $G$ is $\mathcal L=\mbox{span}\{\ket{\psi}: \,  U_s\ket{\psi}=\ket{\psi}\}$. 

Suppose now the system is in a state which is an equal superposition of all elements $g\in G$ acting on $\ket{0}$, i.e., $a(g)= |G|^{-1/2}$ for all $g$. This is obviously a stabilized state because $g\ket{\psi}= |G|^{-1/2} \sum_{g'\in G} gg'\ket{0}=\ket{\psi},\ \forall g\in G$. However, there are states in the stabilized space which are not an equal superposition of all elements $g\in G$ acting on $\ket{0}$. Any superposition of the form 
\be
|G|^{-1/2} \sum_{h\in G', g\in G}a(h)hg\ket{0}
\ee
where $G'$ is a subset of $N$, is still stabilized by $G$.

From now on we will focus on states in the stabilized space that are an equal superposition of the elements of $G$ acting on the reference state $\ket{0}$. The reduced density matrix for an equal superposition state is 
\begin{equation}
\rho_A=|G|^{-1}\sum_{g\in G,\tilde{g}\in G_A}g_A\ket{0_A}\bra{0_A}g_A\tilde{g}_A
\label{rhoa2}
\end{equation}
In this case we obtain an analytical formula for the entropy $S$ depending only on the boundary of the partition $(A,B)$.

Define now the quotient $G/G_B$ and let
\be
f= \frac{|G|}{|G_B|}= \frac{|G|}{d_B}
\label{f}
\ee
be its order. Notice that $f$ is the number of elements in $G$ that act freely on $A$. If there are $l$ independent generators of $G$ acting on $A$, it turns out that $f=2^l$.  Define the group $G_{AB}\equiv G/(G_{A}\cdot G_{B})$. We have $|G_{AB}|= |G|/d_A d_B$. 

In the remaining of this section we generalize to an arbitrary group of spin flips the results presented in \cite{hiz}. We can prove the following result.

{\em Theorem 1.} Consider a partition $(A,B)$ of the spin system, and suppose the system is in an equal superposition of all the group elements $g\in G\subseteq N$, acting on the reference state $\ket{0}$. The entropy of entanglement is $S= \log_2 (f/d_A)= \log_2|G|- \log_2(d_A d_B)= \log_2|G_{AB}|$.

{\em Proof.} We first compute the reduced density matrix $\rho_A$. Consider two elements $g=g_A\otimes g_B$ and $g'=g'_A\otimes g'_B$ in $G$. Then $g'_A= g_A$ if and only if $g'= h g$, with $h\in G_B$, and since $f^{-1}=d_B/|G|$, from equation (\ref{rhoa2}) we obtain
\be
\label{denmat}
\rho_A = f^{-1} \sum_{\substack{g\in G/G_B \\
\tilde{g}\in G_A}} g_A\ket{0_A}\bra{0_A}g_A\tilde{g}_A
\ee
Let us compute the square of the reduced density matrix:
\begin{eqnarray}
\nonumber
\rho_A^2 &=& f^{-2}\sum_{ \substack{g,g'\in
G/G_B\\ \tilde{g},\tilde{g}'\in G_A}}g_A\ket{0_A}\bra{0_A}g_A\tilde{g}_A g'_A\ket{0_A}\bra{0_A}g'_A\tilde{g}'_A\\
\nonumber
&=& f^{-2}\sum_{ \substack{g\in G/G_B\\\tilde{g},\tilde{g}'\in G_A}}g_A\ket{0_A}\bra{0_A}g_A\tilde{g}_A\tilde{g}'_A\\
&=& f^{-2}d_A \sum_{ \substack{g\in G/G_B\\\tilde{g}\in G_A}}g_A\ket{0_A}\bra{0_A}g_A\tilde{g}_A = f^{-1}d_A\rho_A
\end{eqnarray}
Expanding the logarithm in Taylor series we obtain $\log_2\rho_A= \rho_A f d_A^{-1} \log_2(d_A/f)$. Then the entropy of entanglement is 
\begin{eqnarray}
\label{main}
S = \log_2(f d_A^{-1})= \log_2 \frac{|G|}{d_A d_B} = \log_2|G_{AB}|
\end{eqnarray}
concluding the proof. \qed

Notice that if $G=N$, then $N= N_A\cdot N_B$ and the entropy is zero as expected, since in this case the state is an equal superposition of all the basis vectors in the Hilbert space.

Equation (\ref{main}) generalizes the result of Ref.~\cite{hiz} (obtained for the group of star operators in the Kitaev's model \cite{kitaev}) to an arbitrary group $G$ of spin flips. We can interpret equation (\ref{main}) as follows. The state of a spin system contains some information. If we have a bipartition $(A,B)$ of the system, we can consider the information contained {\em exclusively} in $A$ and $B$ as the information contained in the {\em bulk} of the two subsystems. If the system is in a state which is an equal superposition of the elements of a (stabilizer) group $G$ of spin flips acting on $\ket{0}$, the bulk information is contained in $G_A$ and $G_B$. The order of this groups amounts for the ``disorder'' in the bulk of the two subsystems. Then equation (\ref{main}) states that the entropy of entanglement $S$ is given by the difference between the total disorder and the disorder in the bulk. 

Similar results for the entropy of stabilizer states (i.e., for a one-dimensional stabilized space), have been obtained in \cite{fattal}.

\subsection{Spins on a lattice and the entropic boundary law}

\begin{figure}
\putfig{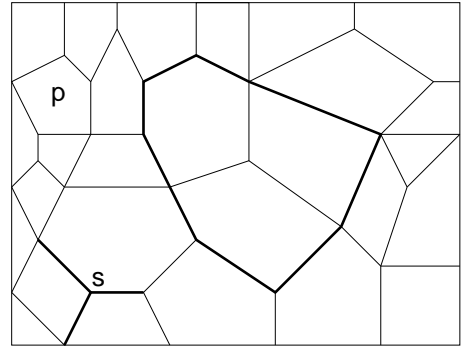}{6}
\caption{A system of spins on an 2D irregular lattice; a typical plaquette and star are denoted by $p$ and $s$, respectively. The subsystem $A$ contains all the spins within the thick boundary. The operators $B_p$ for the plaquettes situated inside the boundary act only on the spins of the subsystem $A$. The $B_p$ of the outside plaquettes that share a link with the boundary act on both subsystems $A$ and $B$ and there is one such plaquette operator for each link in the boundary of $A$. Hence the entropy is $S=L_{\partial A}-n_c$ ($n_c$ is the number of constraints, see text).}
\label{irregular}
\end{figure}

The spin systems hitherto considered have no geometrical structure and hence we have no notion of what the {\em boundary} of the two subsystems is. By giving a lattice structure to the system, we can find a geometrical interpretation of the equation (\ref{main}) as the number of degrees of freedom living on the boundary between the subsystems $A$ and $B$.

Consider an $r$-dimensional lattice with $n$ links and a spin-1/2 attached to each link (the lattice does not need to be regular). Let $n_p$ be the number of plaquettes in the lattice. The Hilbert space is as before $\mathcal{H}=\mathcal{H}_1^{\otimes n}$. Define the stabilizer operators
\begin{equation}
U_p= \prod_{j\in \partial p}\sigma^x_j
\end{equation}
acting on the spins belonging to the boundary of any plaquette $p$. The stabilized space is
\be
\mathcal{L}=\{\ket{\psi}\in \mathcal{H}:\ U_p\ket{\psi}=\ket{\psi}, \forall p\}
\ee

Let $G$ be the group generated by $U_p$, hence $g\mathcal{L}=\mathcal{L}$ for any $g\in G$. If all $U_p$ are independent, then $G$ is generated by the set of all the $n_p$ stabilizer operators. If there are $n_c$ constraints on the set $\{ U_p\}$, then the minimal subset generating $G$ contains $n_p-n_c$ elements and the order of $G$ is $|G|=2^{n_p-n_c}$.

Consider now the stabilized state 
\begin{equation}
\ket{\xi}= |G|^{-1/2}\sum_{g\in G}g\ket{0}
\end{equation}
From Theorem 1, the entropy of entanglement for a state $\ket{\xi}$ corresponding to a partition $(A,B)$ of the lattice is $S = \log_2(f d_A^{-1})= \log_2 (|G|/d_A d_B) = \log_2|G_{AB}|$. For this system $\log_2d_{A(B)}$ is the number of plaquette operators $U_p$ acting exclusively on $A(B)$. Then $S$ is the number $n_{AB}$ of plaquettes acting on both the subsystems $A$ and $B$.

Hence we can give a geometrical interpretation of equation (\ref{main}). For an $r$-dimensional lattice, the entropy is equal to number $n_{AB}$ of degrees of freedom living on the boundary between the two subsystems $A,B$.

\section{The Kitaev's model}\label{kitmod}

\subsection{General formalism}

So far the stabilized states are just some states in a Hilbert space and they do not have any physical meaning. Now we will analyze a case in which the stabilized states are vectors in the ground state manifold of a particular lattice model constructed by Kitaev \cite{kitaev}. This is a 2-dimensional exact solvable spin system on a lattice. Its relevance stems from the fact that it was the first example of the new subject of topological quantum computation \cite{kitaev,top} and because it features {\em topological order} \cite{wen}.

Consider a system of $n$ spins on a (irregular) lattice on a Riemann surface of genus $\mathfrak{g}$. Again, the Hilbert space is $\mathcal{H}=\mathcal{H}_1^{\otimes n}$ and $\dim \mathcal{H}= 2^n$.

The stabilizer operators are the plaquettes 
\begin{equation}
B_p=\prod_{j\in \partial p}\sigma_j^z,
\end{equation}
($j$ labels all the spins belonging to the boundary of a plaquette $p$) and the stars
\begin{equation}
A_s=\prod_{j\in s}\sigma_j^x
\end{equation}
where $j$ labels all the spins sharing a common vertex $s$ (see Fig.\ref{irregular}). On a Riemann surface of genus $\mathfrak{g}$ the number of sites, links (spins) and plaquettes ($n_s,n$ and $n_p$, respectively) obey the Euler's formula: $n_s-n+n_p=2 (1-\mathfrak{g})$. By imposing also $n_s=n_p$, it follows $n_s=n_p=n/2+1-\mathfrak{g}$. We have the following two constraints on the stars and plaquettes:
\begin{equation}
\prod_{\forall s} A_s= \bbbone= \prod_{\forall p} B_p
\end{equation}
so there are only $n_p-1,n_s-1$ independent plaquettes and stars.

Let $G$ be the group generated by the $n-2\mathfrak{g}$ independent stabilizer operators $\{A_s, B_p \}$. We define the protected subspace: 
\begin{equation}
\mathcal{L}=\{\ket{\psi}\in \mathcal{H},\ \ A_s\ket{\psi}=B_p\ket{\psi}=\ket{\psi}\}
\end{equation}
so the states in this set are stabilized by $G$. 

The Hamiltonian of the model is: 
\begin{equation}
H=-\sum_s A_s-\sum_p B_p
\label{hamiltonian}
\end{equation}
The model is exactly solvable because all the stabilizer operators commute with each other (since they share either 0 or 2 links)
\begin{equation}
[A_s,B_p]=0\ , \qquad \forall s,p
\end{equation}
Its ground state is the protected subspace manifold $\mathcal{L}$.

We now show that the ground state manifold $\mathcal{L}$ for a genus-$\mathfrak{g}$ Riemann surface is $2^{2\mathfrak{g}}$-fold degenerate. Since all the stars and plaquettes commute, we can label $n_s+ n_p-2= n-2\mathfrak{g}$ spins out of $n$ and thus the dimension of the ground state is
\begin{equation}
\dim \mathcal{L}= 2^{n-(n-2\mathfrak{g})}= 2^{2\mathfrak{g}}
\end{equation}
hence the system exhibits {\em topological order} \cite{wen}.

\begin{figure}
\putfig{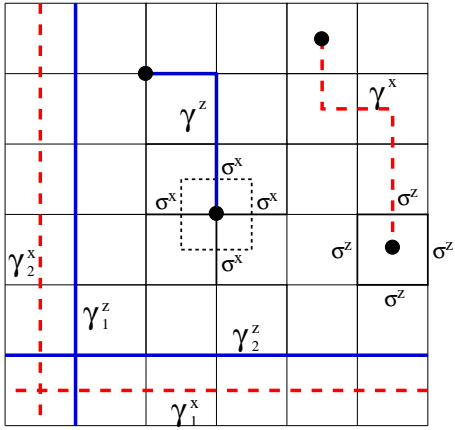}{6}
\caption{(Color online) A $k\times k$ square lattice of the torus; opposite boundaries are identified. The end of an open $z$-string anticommutes with the star based at the site where the string is. Similarly, the end of an open $x$-string anticommutes with the plaquette on the square where it ends. The figure also shows that a star on the lattice corresponds to a plaquette in the dual lattice and vice versa.}
\label{kitaev}
\end{figure}

Another way to see the same thing is to notice that this model features string condensation \cite{wen}. Let $\gamma^z$ ($\gamma^x$) be a curve connecting sites along the links of the lattice (dual lattice) as in Fig.~\ref{kitaev} (for a square lattice). We can define two types of string operators (simply called ``strings''):\\
(i) a {\em z-string} is the product of all $\sigma^z$ operators along the links belonging to a curve $\gamma^z$ running on the lattice;\\
(ii) an {\em x-string} is the product of all $\sigma^x$ operators along the links crossed by a curve $\gamma^x$ (hence running on the dual lattice). The action of an $x$-string is to flip all the spins (i.e., links) intersected by the curve $\gamma^x$.

More formally, a string operator is:
\begin{equation}
W^a[\gamma^a]=\prod_{j \in \gamma^a}\sigma_{j}^{a}\ \ , \qquad a=x,z
\end{equation}
By $j\in\gamma^z$ ($j\in\gamma^x$) we mean all the links belonging to (crossed by) the string $\gamma^z$ ($\gamma^x$). A {\em string-net} is a product of string operators.

Closed strings of both types commute with the Hamiltonian (\ref{hamiltonian}),
\begin{equation}\label{stringcondensation}
[W^a[\overline{\gamma}_a],H]=0, \qquad a=x,z
\end{equation}
where $\overline{\gamma}^a$ is a loop on the lattice or on the dual lattice. This is because a closed string has either $0$ or $2$ links in common with any plaquette or star. However, open strings do not commute with the Hamiltonian. More precisely, open $z$-strings ($x$-strings) anticommute, with star (plaquette) operators with which they share (cross) a single link (see Fig.~\ref{kitaev}).

Equation (\ref{stringcondensation}) implies that there are closed strings in the ground state of the Hamiltonian (\ref{hamiltonian}). We say that we have {\em closed string condensation} \cite{wen} in the sense that closed strings are present in the ground state and that they are not made of smaller pieces that are still present in the ground state. Smaller pieces are indeed open strings and they pay energy. The group $\tilde{G}$ generated by the closed strings is the group of all the closed string-nets of $x$- and $z$-type.

On a genus-$\mathfrak g$ Riemann surface ($\mathfrak g> 0$) there are contractible and non-contractible loops. A loop on the lattice is contractible if is homotopic to the boundary of a plaquette. It turns out that all the string operators based on contractible loops are made of products of stars and plaquettes. This also implies that string operators based on contractible loops have a trivial effect on the ground state $\ket{\xi}\in {\mathcal L}$:
\begin{equation}
W^a[\Gamma]\ket{\xi}= \prod_{j\in\Gamma}\sigma^a_j \ket{\xi}=\ket{\xi}
\end{equation}
where $\Gamma$ is a contractible loop on the lattice (or the dual lattice).

Consider now the non-contractible loops. It is enough to consider only string operators associated to non-contractible loops with winding number $1$. [Since all string operators square to the identity, string operators associated to non-contractible loops with winding number $n$ are equal (modulo a product of stars) to the ones with winding number $n\, \mathrm{mod}\, 2$]. The associated string operators cannot be written in terms of products of star and plaquette operators and hence they have a nontrivial action on the states. But since they still commute with the Hamiltonian, they map ground states into ground states. The algebra ${\bf L}(\mathcal L)$ of linear operators acting on the ground state manifold $\mathcal L$ is the algebra of the closed string operators of $x$- and $z$-type. However, the contractible string operators have a trivial effect on $\mathcal L$, so only the non contractible ones matter. Consider the string operators associated to the non-contractible loops $\{\gamma_i, i=1,...,4\mathfrak{g}\} \equiv \{\gamma^x_1,...,\gamma^x_{2\mathfrak g},\gamma^z_{2\mathfrak g+1},...,\gamma^z_{4\mathfrak g}\}$; the loops $\{\gamma^x_1,...,\gamma^x_{2\mathfrak g} \}$ generate the homotopy group of the Riemann surface. We label the loops such that $\gamma_i^x$ and $\gamma_{i+2\mathfrak g}^z$ intersect, with $i=1,..,2\mathfrak g$, see Fig.\ref{kitaev}:
\begin{eqnarray}\label{w1w2}
w_i\equiv W^{\alpha}[\gamma_{i}] \qquad i=1,...,4\mathfrak g
\end{eqnarray}
where $\alpha=x$ for $i=1,...,2\mathfrak g$ and $z$ otherwise.
We refer to $w_i, i=1,...,2\mathfrak g$ as ``ladder'' operators since they flip all the spins along a ladder going around the non contractible loops of the surface. The pair $\left(w_i,w_{i+2\mathfrak g}\right)$ has the same commutation relations as $(\sigma^x,\sigma^z)$ and generates a 4-dimensional algebra. We see that for any $i=1,...,2\mathfrak g$ we have a copy of the same algebra. Then we have $2\mathfrak g$ mutually commuting copies of the same algebra $\sigma^x,\sigma^z$ and hence ${\bf L}(\mathcal L)$ is $4^{2\mathfrak g}$-dimensional:
\begin{eqnarray}
\nonumber
w_iw_{i+2\mathfrak g}=-w_{i+2\mathfrak g}w_{i},\qquad i=1,...,2\mathfrak g
\\
\left[w_i,w_j\right]=0, \qquad j\ne i\pm 2\mathfrak g,\ i=1,...,4\mathfrak g
\end{eqnarray}
Therefore the ground state manifold $\mathcal L$ is $2^{2\mathfrak g}$-fold degenerate. This degeneracy is the sign of the topological order of this system \cite{wen}, and is robust against arbitrary perturbations \cite{kitaev,wen}. Topological order is the notion needed to describe those states of the matter like fractional quantum Hall liquids \cite{tsui} which are not explained by the Landau theory of symmetry breaking with local order parameters \cite{ginzburg}.

We want now to give an explicit expression for the states in the ground state manifold. Let $\mathcal A$ be the group generated by the stars $A_s$. Let $T$ be the group generated by the $2{\mathfrak g}$ ladder operators of $x$-type $w_i$, with $i=1,...,2\mathfrak g$. Then the elements of $T$ are of the form
\be
w({\bf s})=\prod_{j=1}^{2\mathfrak{g}}w_j^{s_j}
\ee
where ${\bf s}=(s_1,..,s_{2\mathfrak g})$ and $s_j=0,1$, which implies that $|T|= 2^{2\mathfrak g}$. We will call $\overline{N}$ the group of all closed string-nets of the $x$-type:
\be
\overline{N}= \mathcal A \cdot T
\ee

For a generic lattice with $n_s$ stars, $n$ spins and $n_p$ plaquettes, the number of independent star operators is $n_s-1= n- n_p+1- 2\mathfrak g$ and hence the order of $\mathcal A$ is $|\mathcal A|= 2^{n_s-1}= 2^{n-n_p+1-2\mathfrak g}$. The order of $\overline{N}$ is $|\overline{N}|= 2^{n_s-1+ 2\mathfrak g}= 2^{n- n_p+1}$. 

Since on the lattice any loop intersects the boundary of a plaquette in an even number of points, it follows immediately that the states $w({\bf s})\ket{0}$ are stabilized by the plaquettes $B_p$. Then we have $2^{2\mathfrak g}$ states in the stabilized subspace $\mathcal{L}$ given by
\begin{equation}
\ket{\xi({\bf s})}= |\mathcal A|^{-1/2} w({\bf s}) \sum_{g \in\mathcal{A}} g\ket{0}= w({\bf s}) \ket{\xi({\bf 0})}
\end{equation}
which are mutually orthogonal by construction. This shows again that the ground state manifold is $2^{2\mathfrak g}$ degenerate. Each of these states is an equal superposition of the elements in $\mathcal{A}$ and so it falls under the hypothesis of Theorem 1. Notice that an arbitrary superposition of the $\ket{\xi({\bf s})}$'s is still a ground state, but obviously it is not an equal superposition of the elements of a group:
\begin{equation}
\ket{\xi}=\sum_{g \in\mathcal{A},{w(\bf s})\in T} a({\bf s})w({\bf s}) g\ket{0}
\end{equation}
so $\mathcal{L}=\mbox{span}\{\ket{\xi({\bf s})}\}= \mbox{span}\{ T\ket{\xi(\bf 0)} \}$.

\begin{figure}
\putfig{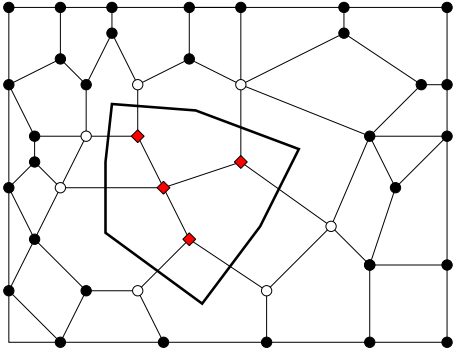}{6}
\caption{(Color online) A lattice region consisting of all the spins inside or crossed by a loop (thick line) on the dual lattice. The $\Sigma_A$ stars based on the sites inside the region $A$ (red diamonds) act exclusively on the subsystem $A$ and the $\Sigma_B$ stars based on the black sites act only on the subsystem $B$. There are $\Sigma_{AB}$ stars (white dots) acting on both subsystems $A$ and $B$. The total number of stars is $n_s= \Sigma_A+ \Sigma_B+ \Sigma_{AB}= n- n_p+2- 2\mathfrak g$, where $n$ ($n_p$) is the number of spins (plaquettes) of the lattice.}
\label{irregular2}
\end{figure}

We now prove that all the basis states $\ket{\xi({\bf s})}$ have the same entanglement.

{\em Proposition 1.} For a given lattice partition $(A,B)$ all the states $\ket{\xi({\bf s})}$ have the same entropy of entanglement $S$.

{\em Proof.} We can decompose the ladder operators as $w({\bf s})\equiv w_A({\bf s}) \otimes w_B({\bf s})$, where $w_{A(B)}({\bf s})$ acts only on the $A$ ($B$) subsystem. From the circular property of the trace and $w^2({\bf s})= \bbbone$, $w^\dag ({\bf s})= w({\bf s})$, ${\bf s}= 0,...,2^{2\mathfrak g}-1$, it follows immediately that all the basis states $\ket{\xi({\bf s})}$ have isospectral reduced density matrices. Since $\ket{\xi({\bf s})}= w({\bf s}) \ket{\xi({\bf 0})}$, then  $\rho_A [\xi({\bf s})] = \mbox{Tr}_B\left(\ket{\xi({\bf s})}\bra{\xi({\bf s})}\right)=\mbox{Tr}_B\left(w({\bf s})\ket{\xi({\bf 0})}\bra{\xi({\bf 0})}w({\bf s})\right)= w_A({\bf s}) \mbox{Tr}_B(\ket{\xi({\bf 0})}\bra{\xi({\bf 0})}) w_A({\bf s})$. Therefore
\be
S(\ket{\xi({\bf s})})= S(\ket{\xi({\bf 0})}),\qquad \forall {\bf s}=0...2^{2\mathfrak g}-1
\ee
and all basis states have the same entropy. \qed

Let us now compute the entropy of entanglement for a state $\ket{\xi({\bf s})}$. The number of closed string-nets acting exclusively on $A(B)$ is $d_{A,B}= 2^{\Sigma_{A,B}}$, where $\Sigma_{A,B}$ is the number of independent star operators acting exclusively on the subsystem $A(B)$. From Theorem 1 we obtain
\begin{eqnarray}
\nonumber
S &=& \log_2 \frac{2^{n_s-1}}{d_A d_B}= n_s-1- \Sigma_A -\Sigma_B \\
&=& n-n_p+1-2\mathfrak g- \Sigma_A -\Sigma_B
\end{eqnarray}
We notice that the {\em topological order} in this model manifests itself in both the degeneracy and the entanglement in the ground state. This suggests the very appealing possibility that entanglement could detect topological order.

The total number of lattice sites is $n_s= \Sigma_A+\Sigma_B+\Sigma_{AB}= n- n_p+2- 2\mathfrak g$, where $\Sigma_{AB}$ is the number of independent star operators acting on {\em both} subsystems $A$ and $B$. We obtain
\be
\label{geomentropy}
S= \Sigma_{AB}-1
\ee
If we choose the partitions in a convenient way, we can give a clear geometrical picture of the formula (\ref{geomentropy}). Let $A$ be the set of all spins inside or crossed by a contractible loop in the dual lattice (see Fig.\ref{irregular2}). The spins intersected by the loop are the {\em boundary} of $A$, while the ones inside are the {\em bulk}. If the loop is {\em convex}, the number of spins $n_L$ in the perimeter $L$ of $A$ is $n_L= \Sigma_{AB}$ (see Fig.\ref{irregular2}). Therefore 
\begin{equation}
\label{loopentropy}
S= n_L-1
\end{equation}

It is interesting that no partition has zero entanglement for all the $\ket{\xi({\bf s})}$ states. The argument is simple: $S=0 \Leftrightarrow |\overline{N}|= d_A d_B \Leftrightarrow \mathcal{A}= \mathcal{A}_{A}\cdot \mathcal{A}_{B}$; but this cannot be satisfied, since there is at least a star or a ladder acting on both $A$ and $B$ for any partition $(A,B)$, hence $S>0$. The group $\mathcal{A}$ splits in $\mathcal{A}=\mathcal{A}_A\cdot\mathcal{A}_B$ for any partition only if is the group generated by the single spin flips, namely $N$. If every spin is shared by at least two generators of $\mathcal{A}$ (which is always the case for star operators on a lattice), then the entropy cannot be zero and we have an absolute entropy.

\section{Ground state entanglement for the Kitaev's model on  a torus square lattice}

In this section we consider a square $k\times k$ lattice on the torus ($\mathfrak{g}= 1$) and we calculate explicitly the entropy $S$ for several bipartitions $(A,B)$ of the lattice. On such a lattice there are $n_s=k^2$ sites and $n=2k^2$ spins.

We have two ladder operators $w_1$ and $w_2$ corresponding to the two non-contractible loops which run along the parallel, and respectively the meridian, of the torus. The group ${\mathcal A}$ generated by the stars has order $|{\mathcal A}|= 2^{n/2-1}$. Then the group $\overline{N}$ is generated by ${\mathcal A}$, $w_1$ and $w_2$ and hence $|\overline{N}|= 2^{n/2+1}$.

The ground state is four-fold degenerate and the vectors $\ket{\xi_{ij}}$, $i,j=0,1$ form a basis, with
\be\label{squarestate}
\ket{\xi_{ij}}=|\mathcal{A}|^{-1/2} w_1^j w_2^i \sum_{g \in {\mathcal A}} g \ket{0}
\ee
An arbitrary vector of the ground state $\ket{\xi}\in \mathcal{L}$ can then be written as
\begin{eqnarray}
\nonumber
\ket{\xi}&=&  \sum_{i,j=0}^1 a_{ij} \ket{\xi_{ij}}\\
\nonumber
&=& |{\mathcal A}|^{-1/2}\sum_{g\in {\mathcal A}} (a_{00}+a_{01}w_1+a_{10}w_2+a_{11}w_1w_2) g\ket{0}\\
&=& |{\mathcal A}|^{-1/2}\sum_{g\in {\mathcal A}} Ug\ket{0}
\end{eqnarray}
where $U\equiv a_{00}\bbbone+ a_{01}w_1+ a_{10}w_2+ a_{11}w_1w_2$ and $\sum_{i,j=0}^1 |a_{ij}|^2= 1$.

The associated density matrix is
\be\label{rhoexpansion}
\rho=\sum_{i,j,l,m=0}^1 a_{ij}\overline{a}_{lm} w_1^j w_2^i \rho_0 w_1^m w_2^l
\ee

Any element $g\in {\mathcal A}$ leaves invariant the ground state, $g \ket{\xi_{ij}}= \ket{\xi_{ij}}$, hence $g \rho_0= \rho_0$, where
\begin{equation}
\rho_0\equiv \ket{\xi_{00}}\bra{\xi_{00}}
\end{equation}

There is another important issue to point out. Suppose we have two ladder operators $w_1$ and $\tilde{w}_1$ with homotopic supports. This means that they are related by an element $g\in {\mathcal A}$, $\tilde{w}_1= g w_1$. Since $g \rho_0= \rho_0$ and $[g, w_1]= 0$, then
\begin{equation}
\tilde{w}_1 \rho_0= g w_1 \rho_0= w_1 \rho_0
\label{wrho}
\end{equation}
Therefore both $w_1$ and $\tilde{w}_1$ have the same effect on $\rho_0$ and hence we can work only with a representative $w_1$.

A proposition useful for computing the entropy in some of the examples below is the specialization of the diagonality condition for the general Kitaev model. We first prove that a generic product of spin flips acting exclusively on the subsystem $A$ commutes with the Hamiltonian only if is a product of closed strings. Let $\overline{N}_A$ be the set of closed string nets that act exclusively on $A$, $\overline{N}_A\equiv\{g\in\overline{N}:g=g_A\otimes\bbbone_B\}$. Then we have the following:

{\em Lemma 2.} Let $g_A \otimes\bbbone_B\in N_A$ be a generic product of spin flips operators acting exclusively on $A$. Then $\left[ H,g_A \otimes\bbbone_B \right]=0 \Leftrightarrow g_A\otimes\bbbone_B\in \overline{N}_A$.

{\em Proof.} The proof is obvious since only closed string nets commute with the Hamiltonian, so if $g_A\otimes\bbbone_B$ commutes with $H$ it must be in $\overline{N}$ and henceforth in $\overline{N}_A$. \qed

{\em Proposition 2.} Suppose the system is in the ground state $\ket{\xi_{00}}$ given by equation (\ref{squarestate}). Then the reduced density matrix $\rho_A$ is diagonal (i.e., $d_A= 1$) if and only if for every $g\in\mathcal A, g_A \not= \bbbone_A$, $g_A$ does not commute with the Hamiltonian: $[g_A, \, H]\ne 0$.

{\em Proof.} ``$\Rightarrow$'': Let $d_A=1$, and suppose, ex absurdo, that there is a non trivial element $g_A\in N$ such that $[g_A,\, H]= 0$. Then from Lemma 2 this means that $g_A\otimes \bbbone_B$ is a nontrivial element of $\mathcal{A}_A$, contradicting the hypothesis.

The reverse implication ``$\Leftarrow$'' obviously holds because for any nontrivial $g_A$ which does not commute with the Hamiltonian, then it follows from Lemma 2 that $g=g_A\otimes\bbbone_B$ is not in $\mathcal A_A$, and thus only the identity belongs to $\mathcal A_A$, which concludes the proof. \qed

A corollary is that if the system is in the ground state $\ket{\xi_{ij}}$, the same proposition holds true by substituting $\mathcal{A}$ with $w_1^jw_2^i\mathcal{A}$. If the system is in a generic ground state $\ket{\xi}$, one just replace $\mathcal A$ with $\overline{N}$.

In the following we compute the entropy for several subsystems $A$ for both $\ket{\xi_{00}}$ and the generic $\ket{\xi}$ ground states. Although the general ground state $\ket{\xi}$ is no longer an equal superposition state and hence we cannot apply Theorem 1, in some cases we can calculate explicitly the von Neumann entropy for $\ket{\xi}$. For completeness we also review some examples given in Ref.~\cite{hiz} for the ground state $\ket{\xi_{00}}$.

\subsection{One spin}\label{1spin}

Take $A$ to be a single spin. In this case it is obvious that no closed string net $g\in \overline{N}$ acts exclusively on $A$, hence $\rho_A$ is diagonal (from Lemma 1) and both eigenvalues are equal to 1/2 (from symmetry, the entries for spin-up and spin-down are equal). Since the entropy is $S= 1$, it follows that any spin is maximally entangled with the rest of the system.

\subsection{Two spins}\label{2spins}

We want now to compute the entanglement between two arbitrary spins of the lattice. To do this, we first obtain the reduced density matrix $\rho_{ij}$ of the two spins by tracing out all the other spins. Since we want to calculate the entanglement {\em between} the two spins, we use as an entanglement measure the concurrence $C$ of the mixed state $\rho_{ij}$ of the two qubits (i.e., the two spins) defined as \cite{wooters}:
\be
C= \max \{0, \sqrt{\lambda_1}-\sqrt{\lambda_2}-\sqrt{\lambda_3}-\sqrt{\lambda_4} \}
\ee
where $\lambda_1,\lambda_2,\lambda_3,\lambda_4$ are the eigenvalues (in decreasing order) of the matrix $\rho_{ij} (\sigma^y \otimes \sigma^y) \rho^*_{ij} (\sigma^y \otimes \sigma^y)$ and $\sigma^y$ is the Pauli matrix. Since no closed string net $g\in \overline{N}$ acts exclusively on the two spins, the reduced density matrix $\rho_{ij}$ is diagonal (again from Lemma 1). Let $\rho_{ij}= \mbox{diag}(a,b,c,d)$. A simple calculation shows that $C= 0$ always, hence there is no two-qubit entanglement between any two spins.

We see that although an arbitrary spin is maximally entangled with the rest of the system, the entanglement is zero between any pair of spins.

\begin{figure}
\putfig{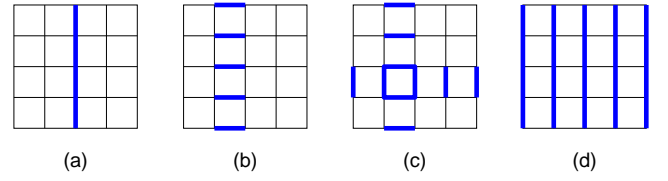}{8.5}
\caption{(Color online) The subsystems $A$ (thick/blue spins) used in calculating the entropy $S$: (a) the spin chain; (b) the vertical ladder; (c) the cross; (d) all the vertical spins.}
\label{partitions}
\end{figure}

\subsection{The spin chain}\label{chain}

 Let $A$ be the set of the $k$ spins belonging to the meridian $\gamma_{z1}$ of the torus as in Fig.\ref{partitions}(a) and consider the system in the state $\ket{\xi_{00}}$. Since the state $\ket{\xi_{00}}$ is the equal superposition of all the group elements in $\mathcal{A}$ acting on $\ket{0}$, we can apply Theorem 1 and the entropy is $S=k^2-1-\log_2d_A-\log_2d_B$. It is obvious that any spin flip on the chain does not commute with the plaquettes sharing that spin, so from the Corollary to Proposition 2 no closed string-net $g\in \overline{N}$ acts exclusively on $A$ and thus $\rho_A$ is always diagonal for any ground state $\ket{\xi}$. In particular, for $\ket{\xi_{00}}$ we have $d_A= 1$. Therefore $S=\log_2 (|G|/d_B)= \log_2 f$ from equation (\ref{f}). The number of possible configurations of spins on the chain $\gamma_{z1}$ is $2^k$, but there are only $f=2^{k-1}$ different configurations of spins in $A$ that enter the ground state, namely the ones with an even number of spin flips. Indeed, we have $k-1$ stars acting independently on the chain and we can obtain all the allowed configurations applying products of these stars (i.e., elements in $\mathcal{A}$ acting freely on the chain), which gives $f=2^{k-1}$. Then the entropy is
\be
S= \log_2 f= k-1
\ee

Let now the system be in a generic $\ket{\xi}$ of the ground state manifold. This state is a superposition with arbitrary coefficients of the four orthogonal states $\ket{\xi_{ij}}$ in $\mathcal{L}$:
\be
\ket{\xi}= \sum_{i,j=0}^1 a_{ij} \ket{\xi_{ij}}= |\mathcal{A}|^{-1/2} \sum_{i,j=0}^1 a_{ij} w_1^j w_2^i \sum_{g \in {\mathcal A}} g \ket{0}
\ee
Since this is not an equal superposition of elements of a group (acting on  $\ket{0}$), we cannot apply Theorem 1. Nonetheless, as shown before, the reduced density matrix for this system is diagonal (in the computational basis). Moreover, all the possible configurations on the chain  are allowed in $\overline{N}\ket{0}$ since all of them can be realized by applying some horizontal ladder $w$. Thus there are $2f= 2^k$ configurations of the chain that are in $\overline{N}\ket{0}$, where $f=2^{k-1}$ is the number of even (odd) spin flips configurations. We see that the states $\ket{\xi_{00}}$ and $\ket{\xi_{10}}$ give states with an even number of spin flips on the chain, while $\ket{\xi_{01}}$ and $\ket{\xi_{11}}$ give states with an odd number of spin flips. The eigenvalues corresponding to the even spin flip configurations are then $|{\mathcal A}| d_B (|a_{00}|^2+|a_{10}|^2)\equiv |{\mathcal A}| d_B \alpha$, while the eigenvalues corresponding to an odd number of spin flips on the chain are $|{\mathcal A}| d_B (|a_{01}|^2+|a_{11}|^2)= |{\mathcal A}| d_B (1-\alpha)$. If $\lambda_i$ are the previous eigenvalues, the entropy is $S= -\sum_{i=1}^{2^k} \lambda_i \log_2 \lambda_i= k-1-\alpha \log_2\alpha -(1-\alpha) \log_2(1-\alpha)$. Defining the binary entropy 
\be
H_2(x)\equiv -x\log_2x- (1-x)\log_2(1-x)
\ee
we obtain
\be
S= k-1+ H_2(\alpha)
\ee

\subsection{The spin ladder}\label{ladder}

 Let $A$ be the set of spins belonging to a ``ladder'', i.e., all the spins intersected by the curve $\gamma_{x1}$, Fig.\ref{partitions}(b). Again, the $A$ partition contains $k$ spins. Consider the system in the state $\ket{\xi_{00}}$. From Proposition 2, no $g \in \mathcal{A}$ acts exclusively on the subsystem $A$, hence $d_A=1$. The number of independent stars acting only on $B$ is $k^2-1- 2k+k$ (there are $k^2-1- 2k$ stars which do not touch the subsystem $A$, plus $k$ {\em pairs} of stars based on the two ends of each spin in $A$, leaving it invariant). Then $d_B= 2^{k^2- k-1}$ and the entropy is
\be
S= k
\ee
This implies immediately that
\be
\mbox{Tr}_B(\rho_0)= 2^{-k} \bbbone_A
\label{rholadder}
\ee
since the $A$-system is in the totally mixed state.

What happens if the system is in a generic ground state $\ket{\xi}$? In general, the reduced density matrix is no longer diagonal (in the computational basis). From equation (\ref{rhoexpansion}) we find
\be
\rho_A=\sum_{i,j,l,m=0}^1 a_{ij}\overline{a}_{lm} \mbox{Tr}_B(w_1^j w_2^i \rho_0 w_1^m w_2^l)
\ee
Since the set $A$ is the vertical ladder, both ladder operators have a particularly simple action: $w_1$ ($w_2$) acts only on subsystem $B$ ($A$). Then $\mbox{Tr}_B(w_2 \rho_0)= w_2 \mbox{Tr}_B(\rho_0)$. Moreover, $\mbox{Tr}_B(w_1 \rho_0)= \sum_{g, g'\in \mathcal{A}} x_A \ket{0_A} \bra{0_A}x_A x_A' \bra{0_B} x'_B w_1 \ket{0_B}$. These scalar products are different from zero if and only if $x'_B=w_1$. This would imply that $x'_A \otimes x'_B=x'_A\otimes w_1$ is a contractible string net in $\mathcal A$ and this can happen only if $x'_A$ is a ladder operator acting fully on $A$, which is impossible (notice that a double ladder is a product of stars and hence a contractible string net). Thus $\mbox{Tr}_B(w_1 \rho_0)=0$. Similarly, $\mbox{Tr}_B(w_1 w_2 \rho_0)= w_2 \mbox{Tr}_B(w_1 \rho_0)= 0$. We also have $\mbox{Tr}_B(w_1 \rho_0 w_1)= \mbox{Tr}_B(\rho_0)$ and $\mbox{Tr}_B(w_2 \rho_0 w_2)= w_2 \mbox{Tr}_B(\rho_0) w_2$. From equation (\ref{rholadder}) we know that for the ladder $\mbox{Tr}_B(\rho_0)= 2^{-k} \bbbone_A$ and we obtain
\begin{equation}
\rho_A= 2^{-k}(\bbbone+ p w_2)
\end{equation}
with $p= 2\,\mbox{Re}(a_{00}\overline{a}_{10}+ a_{01}\overline{a}_{11})$. Since $w_2^2= \bbbone$ and $\mbox{Tr} (w_2)= 0$, the eigenvalues of $w_2$ are $\pm 1$ and they have the same multiplicity, namely $2^{k-1}$. Hence the eigenvalues of $\rho_A$ are $\lambda_\pm = 2^{-k}(1\pm p)$ and the entropy is
\be
S= k- 1 + H_2 \left( \frac{1+p}{2} \right)
\ee

\subsection{The cross}\label{cross}

 The subsystem $A$ includes all the thick (blue) spins in the state Fig.\ref{partitions}(c). This is a system of $2k$ spins. Let the system be in the state $\ket{\xi_{00}}$. Again, no element of $\mathcal{A}$ is able to flip spins only on this subsystem, so $d_A=1$ and the reduced density matrix $\rho_A$ is diagonal (in the computational basis). There are $k^2-1-(2k-1)$ stars acting independently on $B$. The entropy is thus $S= 2k-1$.

\subsection{The vertical spins}\label{vertical}

 We now take $A$ to be the set of all vertical spins of the lattice; then $B$ is the set of all horizontal spins, see Fig.\ref{partitions}(d). The system is considered in the state $\ket{\xi_{00}}$. Since in this case no closed string operator $g\in \mathcal{A}$ acts trivially on either subsystem, we have $d_A= d_B=1$. The entanglement is $S= k^2-1$, which is the maximum possible value for a $\ket{\xi_{ij}}$ state.

\begin{figure}
\putfig{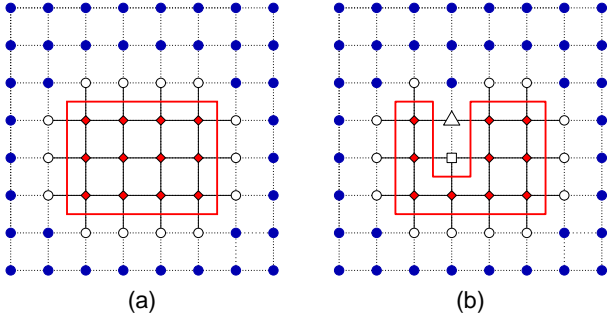}{8}
\caption{(Color online) A region $A$ of the lattice obtained by taking all the spins (thin/black lines) inside or crossed by a loop (thick/red line). $\Sigma_{A,B}$ are the number of sites whose stars act exclusively on $A$ (diamonds/red sites) and respectively, $B$ (solid/blue circles); $d_{A,B}= 2^{\Sigma_{A,B}}$. The number of sites with stars acting on both subsystems (open/white sites) is $\Sigma_{AB}= n_1+n_2+n_3$; $n_i$ is the number of (open/white) sites having $i$ nearest neighbours inside $A$. Area conservation implies $\Sigma_A+\Sigma_B+\Sigma_{AB}= k^2$. The entropy is $S= \Sigma_{AB}-1$. (a) If the boundary is a convex loop (i.e., a rectangle), $\Sigma_{AB}$ is equal to the boundary length $L_{\partial A}$ (in lattice units), since in this case $n_2=n_3=0$; the entropy is $S= L_{\partial A}- 1$. (b) If the boundary is an arbitrary loop, $L_{\partial A}= n_1+2n_2+ 3n_3$, hence the entropy is $S= L_{\partial A}-n_2-2n_3- 1$; in the figure $n_2=1$ (open/white triangle) and $n_3=1$ (open/white square).}
\label{bulk}
\end{figure}

\subsection{The disk}\label{disk}

In this section we take the system $A$ to be a disk, i.e., a region homeomorphic to a plaquette. Let us start, as usual, by assuming the system is in the state $\ket{\xi_{00}}$. Consider a contractible loop $\gamma$ on the dual lattice and let $A$ be the set of spins inside or intersected by $\gamma$ (Fig.~\ref{bulk}). Let $\Sigma_{A,B}$ be the number of sites whose stars act only on $A$, respectively $B$; $\Sigma_{A,B}$ is the area of $A,B$ in lattice units. Let $\Sigma_{AB}$ be the number of sites with stars acting on both subsystems. We also have $d_{A,B}= 2^{\Sigma_{A,B}}$. Area conservation implies $\Sigma_A+ \Sigma_B+ \Sigma_{AB}= k^2$. The entropy is $S= k^2-1-\Sigma_A-\Sigma_B= \Sigma_{AB}-1$. Let $n_i$, $i=1...3$, be the number of sites in $\Sigma_{AB}$ having $i$ nearest neighbours inside $A$. Then
\be
\Sigma_{AB}= n_1+ n_2 + n_3
\ee
and the boundary length is
\be
L_{\partial A}= n_1+ 2n_2 + 3n_3
\label{perimeter}
\ee
If $\gamma$ is a convex loop (Fig.~\ref{bulk}a), then $\Sigma_{AB}= L_{\partial A}$ (since $n_2=n_3=0)$ and the entropy is equal to the perimeter of the boundary (up to a constant) $S= L_{\partial A}-1$. In general, if the boundary of $A$ is an arbitrary loop $\gamma$ on the lattice (Fig.~\ref{bulk}b), the entropy is
\be
S= L_{\partial A}- n_2- 2n_3- 1= \frac{1}{3} (L_{\partial A}+ 2n_1 + n_2)- 1
\label{S}
\ee
We see that for the disk $S$ is exactly the {\em geometric entropy} \cite{callan} of a spatial region $A$.

It is easy to prove that the entropy is bounded from above and below by two linear functions. From eq.~(\ref{perimeter}) it follows that $n_i \le L_{\partial A}/i$ and $2n_2+ 3n_3 \le L_{\partial A}$. Then eq.~(\ref{S}) implies
\be
\frac{1}{3}L_{\partial A} -1 \le S\le \frac{7}{6}L_{\partial A}-1
\ee

Let us now consider the system in the generic ground state $\ket{\xi}$. The expansion of the reduced density matrix is $\rho_A=\sum_{i,j,l,m=0}^1 a_{ij}\overline{a}_{lm} \mbox{Tr}_B(w_1^j w_2^i \rho_0 w_1^m w_2^l)$. From equation (\ref{wrho}), for the disk we can choose the ladders $w_{1,2}$ such that they act only on $B$, hence $w_{iA}= \bbbone_A$, $i=1,2$. Then it follows immediately that $\mbox{Tr}_B(w_i\rho_0 w_i)= \mbox{Tr}_B(\rho_0), i=1,2$, and $ \mbox{Tr}_B(w_1 w_2 \rho_0 w_1 w_2)=\mbox{Tr}_B(\rho_0)$. As in section \ref{ladder}, $\mbox{Tr}_B(\rho_0 w_1)= 0$, since the ladder $w_1$ cannot act exclusively on subsystem $A$. A similar reasoning implies also $\mbox{Tr}_B(\rho_0 w_2)= 0= \mbox{Tr}_B(\rho_0 w_1 w_2)$ and we obtain $\rho_A= \mbox{Tr}_B(\rho_0)$; hence the entropy is the same as in the previous case, $S= L_{\partial A}- n_2- 2n_3- 1$. Thus for the disk the entropy obeys the boundary law for any ground state $\ket{\xi}$.

\begin{table}[t]
\caption{The entropy $S$ of the systems analyzed in text for two ground states, $\ket{\xi_{00}}$ and the generic $\ket{\xi}$; for two spins the value shown is the concurrence $C$. The constants are $\alpha= |a_{00}|^2+|a_{10}|^2$ and $p= 2\,\mbox{Re}(a_{00}\overline{a}_{10}+ a_{01}\overline{a}_{11})$; $H_2(x)= -x\log_2x -(1-x)\log_2(1-x)$ is the binary entropy.}
\begin{ruledtabular}
\begin{tabular}{l|cr}
 & $\ket{\xi_{00}}$ & $\ket{\xi}$ \\
\hline
0. one spin & 1 & 1 \\
1. two spins $i,j$ & $C=0$ & $C=0$ \\
2. spin chain & $k-1$ & $k-1+ H_2(\alpha)$ \\
3. spin ladder & $k$ & $k-1+ H_2(\frac{1+p}{2})$ \\
4. cross & $2k-1$ & -- \\
5. vertical spins & $k^2-1$ & -- \\
6. the disk & $L_{\partial A}- n_2- 2n_3- 1$ & $L_{\partial A}- n_2- 2n_3- 1$ \\
\end{tabular}
\end{ruledtabular}
\label{tab1}
\end{table}

\section{Conclusions}

Apart from being one of the most striking conceptual features of quantum mechanics, entanglement proves also to be a powerful tool in the study of many body spin systems, as several articles pointed out recently.

The first topic to which our article is related is the study of entanglement in spin systems. Several authors have calculated the entanglement in 1D spin chains. In the case of $XY$ and Heisenberg models, the authors in Refs.~\cite{latorre,cirac} calculated the entanglement between a spin block of size $L$ and the rest of the chain. They found two characteristic behaviors. For critical spin chains, the entanglement scales like $S\sim \log_2 L$, whereas for the noncritical case $S$ saturates with the size $L$ of the block. This result is in agreement with the result for black-hole thermodynamics in 1+1 dimensions \cite{larsen,preskill}, which suggested a connection between the entanglement measured in quantum information and the entropy of the vacuum in quantum field theories.

In this article we investigated bipartite entanglement in spin systems for states in a stabilized space. For a bipartite system in a pure state, the von Neumann entropy of the reduced density matrix is the unique measure characterizing the entanglement between the two subsystems. We showed that for states that are an equal superposition of all the elements of a stabilizer group generated by spin flips, the entanglement entropy of a bipartition $(A,B)$ depends only on the degrees of freedom belonging to the boundary between the two subsystems. This property provides an interesting link to the holographic principle. As an example we studied the entanglement present in the ground state of the Kitaev's model. Apart from its special interest in quantum computation (it was the first example of topological quantum computing), this model is also relevant {\em per se}, due to the nontrivial topology and to the specific nature of the spin-spin interaction which generates topological order. On a Riemann surface of genus $\mathfrak{g}$ the degeneracy of the ground state is $4^{\mathfrak{g}}$ and it is stable against local perturbations. We found analytical results for the ground state entropy of several bipartitions $(A,B)$ of a toroidal square lattice. In this case, although no two spins of the lattice are entangled (the concurrence is zero for any pair of spins), the ground state has genuine multi-body entanglement. For a convex region $A$ of the lattice, its geometric entropy is linear in the length of the boundary. Moreover, for states which are an equal superposition of all elements $g\in G\subset N$ acting on $\ket{0}$, no partition has zero entanglement, so the system has an {\em absolute entanglement}. Finally, we argued that entanglement can probe the topology of the system and raised the very interesting question of whether it could detect (or measure) topological order. 

It is relevant to put our results in perspective and to compare them with known results. The {\em holographic principle} (HP) emerged recently as a paradigmatic universal law \cite{holographic}. A simplified statement of HP is: {\em The maximum entropy of a region is proportional to the area of its boundary}. It apparently contradicts the na\"\i ve expectation that the entropy of a region should be proportional to its volume. The entropic area law appears as a recurrent pattern and has been recovered in several (apparently unrelated) physical systems. In black hole thermodynamics it is expressed as the Beckenstein-Hawking law, $S_{BH}= A/4$: the entropy of a black hole is a quarter of its horizon area (in Planck units) \cite{bh}. For a scalar field in 2+1 and 3+1 dimensions, Srednicki \cite{srednicki} showed that the entropy of a region $R$ is proportional to the area of its boundary, and not to its volume. Recently Plenio {\em et al.} \cite{plenio} found analytically the same behavior for the entropy in the case of a harmonic lattice system in $d$-dimensions.

The entropic boundary law recovered in this article for spin systems provides another instance of the universality of HP. Due to the close relationship between concepts like entropy and entanglement, the holographic principle gives insights into fundamental questions of quantum information theory, like {\em What is maximum information content needed to describe a region $R$?} or {\em How much information can be stored inside a system A?} This confluence of diverse fields, like black hole thermodynamics, QIT, and spin systems, can bring together insights and shed new light on fundamental problems.

P.Z.~gratefully acknowledges funding by European Union project TOPQIP (contract IST-2001-39215).


\end{document}